\documentclass[nonacm, sigplan]{acmart}
\AtBeginDocument{%
  \providecommand\BibTeX{{%
    \normalfont B\kern-0.5em{\scshape i\kern-0.25em b}\kern-0.8em\TeX}}}

\usepackage{adjustbox}

\usepackage{algorithmic}
\usepackage{comment}
\usepackage{caption}
\usepackage{subcaption}
\usepackage{graphicx}
\usepackage{textcomp}
\usepackage[htt]{hyphenat}
\usepackage{hyperref}
\usepackage{xurl}
\hypersetup{breaklinks=true}
\usepackage[noabbrev,capitalize]{cleveref}
\usepackage{xcolor}
\usepackage{glossaries}
\usepackage{xspace}
\usepackage[inline]{enumitem}
\usepackage{tabularx}
\newcommand*{\ie}{\textit{i.e.,}\@\xspace}
\newcommand*{\eg}{\textit{e.g.,}\@\xspace}
\newcommand*{\cf}{\textit{cf.}\@\xspace}

\newacronym{ai}{AI}{Artificial Intelligence}
\newacronym{adas}{ADAS}{Advanced Driver Assistance System}
\newacronym{als}{ALS}{Adaptive Light System}
\newacronym{bert}{BERT}{Bidirectional Encoder Representations from Transformers}

\newacronym{cad}{CAD}{Computer Aided Design}
\def\cad{\gls{CAD}\xspace}

\newacronym{CPS}{CPS}{Cyber-Physical System}

\def\cpss{\glspl{CPS}\xspace}

\newacronym{NLP}{NLP}{Natural Language Processing}
\def\nlp{\gls{NLP}\xspace}

\newacronym{dsl}{DSL}{Domain Specific Language}
\def\dsl{\gls{dsl}\xspace}
\def\dsls{\glspl{dsl}\xspace}

\newacronym{gpt}{GPT}{Generative Pre-trained Transformer}

\newacronym{MDRE}{MDRE}{Model-Driven Requirements Engineering}
\def\mdre{\gls{MDRE}\xspace}

\newacronym{mda}{MDA}{Model-Driven Architecture}

\newacronym{nsp}{NSP}{Next Sentence Prediction}

\newacronym{cpu}{CPU}{Central Processing Unit}

\newacronym{lstm}{LSTM}{long short-term memory}
\newacronym{nlp}{NLP}{natural language processing}

\newacronym{SD}{SD}{Synchronous Drive}

\newacronym{SI}{SI}{International System of Units (Système International d'unités)}

\newacronym{OMG}{OMG}{Object Management Group}

\newacronym{V\string&V}{V\string&V}{Validation and Verification}

\newacronym{BPMN}{BPMN}{Business Process Model and Notation}

\newacronym{MDD}{MDD}{Model Driven Development}

\newacronym{ERS}{ERS}{Entity Relationship Schemata}

\newacronym{OCL}{OCL}{Object Constraint Language}

\newacronym{UMAP}{UMAP}{Uniform Manifold Approximation and Projection}

\newacronym{UML}{UML}{Unified Modeling Language}
\def\uml{\gls{UML}\xspace}

\newacronym{UML/P}{UML/P}{\gls{UML}/P}

\newacronym{CD}{CD}{Class Diagram}

\newacronym{OD}{OD}{Object Diagram}

\newacronym{AD}{AD}{Activity Diagram}

\newacronym{OM}{OM}{Object Model}

\newacronym{MBT}{MBT}{Model-Based Testing}

\newacronym{AST}{AST}{Abstract Syntax Tree}

\newacronym{SMT}{SMT}{Satisfiability Modulo Theories}

\newacronym{SE}{SE}{Software Engineering}

\newacronym{PDP}{PDP}{Product Development Process}

\newacronym{OCL/P}{OCL/P}{OCL/Programmable}

\newacronym{sysml}{SysML}{Systems Modeling Language}

\newacronym{xml}{XML}{Extensible Markup Language}

\newacronym{ad}{AD}{Activity Diagram}

\newacronym{sc}{SC}{Statechart}

\newacronym{ucd}{UCD}{Use Case Diagram}

\newacronym{uc}{UC}{Use Case}

\newacronym{sd}{SD}{Sequence Diagram}

\newacronym{bdd}{BDD}{Block Definition Diagram}

\newacronym{ibd}{IBD}{Internal Block Diagram}

\newacronym{mde}{MDE}{Model-Driven Engineering}
\def\mde{\gls{mde}\xspace}

\newacronym{CAD}{CAD}{Computer-Aided Design}

\begin{document}

\title{Technical Report on Neural Language Models and Few-Shot Learning for Systematic Requirements Processing in MDSE}

\author{Vincent Bertram, Miriam Boß, Evgeny Kusmenko, Imke Helene Nachmann, Bernhard Rumpe, Danilo Trotta, Louis Wachtmeister}
\email{{bertram,boss,kusmenko,nachmann,rumpe,wachtmeister}@se-rwth.de}
\affiliation{%
  \institution{RWTH Aachen University, Software Engineering}
  \streetaddress{Ahornstr. 55}
  \city{Aachen}
  \country{Germany}
  \postcode{52074}
}

\begin{abstract}
Systems engineering, in particular in the automotive domain, needs to cope with the massively increasing numbers of requirements that arise during the development process. To guarantee a high product quality and make sure that functional safety standards such as ISO26262 are fulfilled, the exploitation of potentials of model-driven systems engineering in the form of automatic analyses, consistency checks, and tracing mechanisms is indispensable. However, the language in which requirements are written, and the tools needed to operate on them, are highly individual and require domain-specific tailoring. This hinders automated processing of requirements as well as the linking of requirements to models. Introducing formal requirement notations in existing projects leads to the challenge of translating masses of requirements and process changes on the one hand and to the necessity of the corresponding training for the requirements engineers.

In this paper, based on the analysis of an open-source set of automotive requirements, we derive domain-specific language constructs helping us to avoid ambiguities in requirements and increase the level of formality. The main contribution is the adoption and evaluation of few-shot learning with large pretrained language models for the automated translation of informal requirements to structured languages such as a requirement DSL. We show that support sets of less than ten translation examples can suffice to few-shot train a language model to incorporate keywords and implement syntactic rules into informal natural language requirements.
\end{abstract}

\settopmatter{printfolios=true} 
\maketitle

\section{Introduction}

\label{sec:Introduction}

Innovation of modern systems comes from realizing complex functionalities through the interaction of software, electrical and mechanical components. \Glspl{adas},
autonomous vehicles, and other \cpss have emerged from this ongoing trend.  Due to the heterogeneous and interactive nature of these systems-of-systems, their engineering has become a highly effortful task that faces many challenges~\cite{Tyagi2021,KRRW17,KRS+18a,KKR+18,KKR19}. 
One of them is certainly the need for collaboration of experts from different domains~\cite{DRW+20}. 
This challenge also manifests in the rising number of requirements that address stakeholders from heterogeneous domains.
In systems engineering, and in particular the automotive domain, requirements are captured as documents that contain text mainly in natural language~\cite{Liebel19} often with additional information provided through, \eg pictures or \cad models. 
Experts interpret these textual requirements to enter the design phase, and most often derive details of the implementation directly from them~\cite{FR07}. 
The ambiguity of natural language, in particular, when interpreted by experts from different backgrounds, 
as well as the increasing number of requirements may result in decreasing product quality, system failures which are currently detected at late development stages~\cite{DGH+19} and hinders the implementation of functional safety standards such ISO 26262. 
Furthermore, the document-based approach to requirements engineering prevents agile development, where automated analyses and syntheses should enable early error detection and fast feedback for the developers.

What is needed are tools to capture, analyze, and process requirements systematically during all phases of the development cycle. An approach to achieve this is \mde~\cite{FR07}, which utilizes models as the primary development artifacts. 
That is, these models serve as documentation and communication basis for engineers, but also as input for analyses and syntheses, such as verification~\cite{KPRR20a}, test case~\cite{DGH+19}, or code generation~\cite{DGM+21}. For instance, \mde can be applied to facilitate the design of \gls{ai}-based systems \cite{KNP+19,GKR19,KPRS19,AKK+21}. Approaches to introduce \mde in the automotive requirements engineering exist~\cite{Loniewski10}, but
introducing \mde comes with initial costs and efforts for training the domain experts in modeling, and most often, for translating many documents to models~\cite{Buchhiarone2020}.
An advantage of using \dsls rather than general purpose modeling languages such as the \uml is that their syntax and semantics~\cite{HR04} can be designed to be  intuitive for the model users. 
As requirements are captured in natural language, we assume that a textual \dsl that offers sentence structures and wording that is close to the current formulation of requirements increases intuitiveness of both usage and understanding of models in this \dsl significantly. 
However, in addition to the \gls{dsl} development costs and the \gls{dsl} training, once the \gls{dsl} is developed, the translation of old, unstructured requirements to models in the \dsl can be a tremendous effort due to the high number of requirements, requiring time and modeling know-how from the translating developer. 

In this paper, we analyze an open-source set of automotive requirements for \gls{adas} and \gls{als} to understand
where formulation inaccuracies occur and how targeted \gls{dsl} constructs can help eliminate these inaccuracies and increase the level of formality and consistency in these requirements. \textbf{The goal and main contribution of this paper is the application and evaluation of few-shot learning of large neural natural language models for the translation of given unstructured requirements to sentences incorporating the new formal \gls{dsl} constructs.} 

Such translation models can be used 1) during the introduction phase of a \gls{dsl} to automatically translate existing or legacy natural language requirements into the new \gls{dsl} syntax and 2) to correct natural language inputs in a smart editor when a requirement engineer writes a new requirement as natural text. With this automation supported by the fact that few-shot learning requires only a handful of translation examples to learn a given translation task, \textbf{our aim is to facilitate the introduction of highly specialized requirement \glspl{dsl}, e.g., targeting a single department of a company using specific wording or even a single project.}

The remainder of this paper is structured as follows.
\Cref{sec:Preliminaries} introduces the technical foundations of our approach; \Cref{sec:MDRE} highlights the challenges and potentials of an \mde approach for requirements engineering within engineering domains, driven by natural-language text-based documents. 
\Cref{sec:RelatedWork} outlines related work in this area. 
\Cref{sec:ReqDSL} presents an example \dsl for capturing requirements in the automotive domain. 
\Cref{sec:Translation} details the automatic translation from natural-language to the \dsl. 
In \Cref{sec:Evaluation} we evaluate the approach in multiple experiments. We discuss threats to validity in \Cref{sec:threats} before \Cref{sec:Conclusion} concludes the paper. This paper is an extended version of the corresponding SLE publication by Bertram et al. 2022 \cite{BBK+22a}.

\section{Preliminaries}
\label{sec:Preliminaries}
\subsection{Neural Language Models}

\begin{sloppypar} 
Our approach for text-to-\dsl translation of requirements relies heavily on large transformer-based neural language models. The original transformer architecture \cite{VSP+17} is a groundbreaking neural network architecture for sequence-to-sequence processing. It has an encoder-decoder structure, where the encoder receives
a sequence as its input and encodes its content to pass it to the decoder. The decoder iteratively creates the target sequence using a graph search method, e.g., beam search. Instead of using recurrent neurons such as \glspl{lstm} the transformer architecture relies exclusively on the attention mechanism to grasp dependencies
between sequence elements.
\end{sloppypar}

While we are not going to use the original transformer architecture itself, all models employed in this paper are its derivatives.
For the automatic translation from natural language requirements to the \dsl, we utilize a derivative of \gls{gpt}~\cite{RNSS18}, which is tailored towards text generation. GPT is a transformer-based decoder-only language model that employs a semi-supervised learning approach~\cite{RNSS18}. 
The authors showed that generative unsupervised pretraining on unlabeled data, where, given a sequence of words, the network is supposed to learn to predict the next word with the highest likelihood, and subsequent supervised fine-tuning of the pretrained parameters for a specific downstream task outperformed discriminatively trained models. \gls{gpt} language models have evolved over the last few years and various variants exist.

In \cite{RWC+19}, the authors showed that, given a sufficiently large capacity and a large, varied text corpus in training, a language model can solve tasks across different domains for which it has not experienced explicit training, also referred to as zero-shot learning. Furthermore, in \cite{BMR+20}, the authors show that zero-shot learning is outperformed by few-shot learning. 
In few-shot learning, the model is given a \textit{support set} consisting of a very small number of training examples demonstrating how to solve a new task as part of the model's input. No weight updates are necessary, i.e. no classical training is performed. The support set is input into the model as part of the query. Based on this context, the language model can then solve the new task for a new input. In contrast to zero-shot learning, few-shot learning enables targeted training for very specific tasks, making it 
particularly interesting for requirements engineering, a discipline heavily relying on natural language and where training data is often scarce. 

For the automatic translation of natural language requirements to a model in the \dsl we mainly rely on GPT-J-6B~\cite{mesh-transformer-jax,gpt-j}, an open-source language model based on the 6.7B GPT-3~\cite{BMR+20} network and its hyperparameters. Similarly to GPT-Neo \cite{gpt-neo}, it is also trained on the Pile dataset~\cite{GBB+20}. According to the authors, its performance is almost on par with the 6.7B GPT-3 network, and it is the best-performing publicly available transformer language model in terms of zero-shot performance on various down-streaming tasks\footnote{https://arankomatsuzaki.wordpress.com/2021/06/04/gpt-j/}.

\subsection{Datasets}
The method presented in this paper is evaluated on a publicly available dataset published by Daimler AG, fostering reproducibility. The dataset stems from the automotive domain and consists of 120 textual requirements~\cite{BMR+17}. It contains natural language requirements for two typical automotive systems, namely \gls{als} and \gls{adas}. 

The requirements of the \gls{als} describe a set of system functions: The functionality that causes the vehicle's direction indicators to flash in response to the steering column lever and hazard warning flasher switch.  A function to lower the beams depending on the rotary light switch position and the vehicle setting for daytime running light. An adaptive high beam to control the high beam headlamp depending on the high beam switch and the detection of oncoming vehicles.

For the \gls{adas} system, the dataset contains requirements concerning the main components for adaptive cruise control which maintains the distance to the vehicle in front and a speed set either manually by the driver or via traffic sign detection, provides a distance warning and an emergency brake assistant which reacts to stationary obstacles and to moving obstacles.

The requirements address a variety of typical features of current embedded software-intensive systems, (i.e., software is the key element to realize the functionality) from the automotive domain such as highly distributed and partially safety-critical functionality, hard real-time requirements, a combination of mechanical, mechatronic and electronic components both having reactive and regulating functional behavior and numerous product variants (e.g., due to of different product architectures, legal requirements or special equipment).

\section{Model-Driven Requirements Engineering: Challenges and Potentials}
\label{sec:MDRE}
As the number of requirements increases with the growing complexity of \cpss, preventing inconsistencies and redundancies has become a highly effortful and error-prone task when performed manually. 
\mde holds the promise to mend these issues by utilizing modeling languages to formulate requirements in a way that they become comprehensive both for computers and humans within the development process~\cite{Buchhiarone2020,Cabot2018,KKR+18}. 
This section outlines the challenges and potentials of introducing \mdre in the automotive domain and defines a set of requirements for requirement \dsls to overcome these challenges.

In \mde models are the primary source of information as their mathematical semantics and strict syntax enable a unique understanding of the modeled subject among all stakeholders and computers. 
Thereby, automated analyses and syntheses performed by the latter become meaningful and therefore practically useful. However, introducing \mde faces major challenges, which~\cite{Cabot2018} summarizes by stating that the benefits of \mde do not outweigh its costs, which is also the case for \mdre. 
Part of the arising costs stem from the facts that (1) training engineers to use the modeling languages correctly and efficiently is effortful~\cite{Hutchinson2014}, and (2) transferring the existing documents to models is very time-consuming~\cite{FR07}, since modeling itself is an effortful and tedious task~\cite{Pati2017}.

Language engineering means the process of creating suitable \dsls that serve as modeling languages within \mde of a specific domain \cite{CFJ+16}. 
Naturally, if this process aims at creating a language that is intuitive for the experts (who do not have a computer-science background), the training effort decreases and with that, the cost of introducing \mde, tackling challenge (1).
The contributions of this paper towards tackling these challenges rely on the idea that a textual \dsl whose syntax is very close to the phrasing of requirements in natural language used by the experts is intuitive to these experts by construction. 

However, engineering such a \dsl requires extensive analyses of the natural language used to formulate these requirements, which increases the cost of language engineering significantly~\cite{FR07} and would probably outweigh the cost saving achieved by intuitiveness.
Further, given a \dsl, automating (parts of) the process to translate natural language requirements to models addresses challenge (2).
An implementation of such an automated translation is most likely based on machine learning and \nlp which, to be useful, requires an extensive set of labeled training data. 
Creating the latter, again, increases the costs of implementing the translation in a way that would probably outweigh its benefits. 
However, we assume, that the closer the phrasing of the output (models in the \dsl) is to the phrasing of the input (requirements in natural language), the more likely it is that few-shot learning approaches suffice for the training. 
These approaches only need small amounts of labeled data for effective training, and therefore reduce the costs to create a sufficient training dataset.
Further, requirements engineering is a very broad field in automotive systems engineering that involves experts from various backgrounds. 
Engineering a \dsl to capture requirements in this domain will be an iterative process during which the language will evolve along with the tools to interpret, analyze and process the models, and this evolution will continue throughout the maintenance phase. 
A modular design of the language and the underlying tooling is therefore crucial~\cite{BPR+20,BEH+20}. 

We derived the following set of requirements on \dsls that facilitate introducing \mdre in the automotive domain by reducing the training and translation efforts: 
\begin{itemize}
        \item[\textbf{R1:}] Since the modelers will not necessarily have a computer-science background, the \dsl's syntax shall be based on natural language.
        \item[\textbf{R2:}] To make the language as intuitive as possible for its users and to enable the application of few-shot learning to implement an automatic translation from natural language requirements to models, the \dsl's syntax shall be as close to the phrasing of requirements in natural language used by the modelers as possible.
        \item[\textbf{R3:}] In the \dsl, requirements shall be formulated consistently and with a precise meaning understood by relevant stakeholders, enabling automatic interpretation.
        \item[\textbf{R4:}] The language and its tooling shall be extensible and maintainable by a modular design.   
\end{itemize}

\section{Related Work}
\label{sec:RelatedWork}
\subsection{Formalization of Requirements}

\paragraph{Requirement \dsls }
Trigger action patterns or If-Then constructs as we are going to use them in our \dsl are widely used in industry. One example is the graphical formalization language Simplified Universal Pattern~\cite{TBH16}, which has shown that a large part of the requirements from the automotive domain can be formalized with this concept~\cite{BBB+18}.
An approach that utilizes a \dsl for automating type checking of requirements for detecting erroneous syntactic phrasings automatically is proposed in~\cite{MSL15}.
Therein, the \dsl is defined by a context-free grammar and an ontology that defines axiomatic terms and types to allow including typed terms within the requirements. 
Further tooling includes automatic consistency checking~\cite{MSL16}.
In~\cite{KC05}, and~\cite{YCC15} the authors propose approaches to verify the consistency of requirements using temporal logic.
Very similar to our \dsl, these approaches use a structured English which is defined through a grammar to enable automatic logical analyses on requirements written by non-computer scientists.
However, these approaches do not aim at minimizing the difference between the sentence constructs offered by the \dsl and the already established natural language used for requirements engineering, which will increase intuitiveness and enable few-shot learning for automated translation.

\paragraph{Structuring Natural Language through Sentence Patterns}
In addition to \dsls for requirements development, requirements templates are widely used for standardizing the wording of requirements and increasing their quality. A requirements template provides the requirements engineer with a set of sentence templates as a guide for requirements development. In German-speaking countries, for example, the MASTER template based on \cite{PR21} is widely used. In addition to the experience-based templates, in the literature there exist approaches that deal with the extraction of requirements templates from legacy requirements. For example, \cite{EAF16} develops a set of sentence templates for quality requirements, \cite{FPQ+10} presents a metamodel for software requirements. In addition, the PABRE framework presents a method for using requirements templates \cite{FQR+13} based on a catalog of 29 QR patterns \cite{RFQ+09} and 37 non-functional patterns \cite{PQF+12}.
Because requirements templates provide only a guideline to write requirements, there are no means to detect deviations from the wording proposed by the templates. This has the advantage that requirement engineers and template designers are not limited to formally defined sentence structures when writing requirements. Hence, this approach has the major disadvantage that implementations of automated requirements processing cannot assume correct phrasing of the input.

\paragraph{Formalization by Applying Logic}
Approaches that formalize natural language requirements by translations into formal languages that allow interpretations based on logic have been researched for decades, and we will just name a few interesting examples here:
The approach presented in~\cite{Gervasi2005} translates natural language requirements into a representation in deontic logic which allows to check for consistency. 
The approach does not utilize machine learning for the translation, but relies on ``typographical adjustments, tokenization and morphosyntactic analysis'' to extract  buckets, \ie pieces of a specific kind of information, from the sentences. 
Similarly, \cite{Dalianis1992} propose a technique that enables users to query a conceptual model of the system in natural language and retrieve answers in natural language. 
The approach targets the verification of natural language requirements given a formal model of the system, \ie a model in a \dsl. 
Another concept for verifying requirements is proposed in~\cite{Goessler2009} which perceives requirements and a component design each as contracts and verify their correctness by a conjunction of both contracts. 
The technique relies on modal logic. 
An overview of the research conducted in formalizing legal requirements is given in~\cite{Otto2007}.
A very recent approach~\cite{Zaki2021} utilizes \nlp to transform natural language requirements into a requirements capturing model, which is a semi-formal model. 
The approach introduces automatic transformations from this representation into metric temporal logic and computation tree logic. 
All of these approaches take a requirement in natural language as input, and transform it into a structured format that is comparable to a model in a \dsl before this model is the input for an automated processing. 
These approaches, however, do not assure that engineers formulate requirements such that they are uniquely understood among the human stakeholders as well: The intermediate model remains internal to the automated processing, and, being a rephrasing as logical formulas, has an unintuitive format for most human readers. 
 
Representing requirements in a textual \dsl structures the sentences such that their understanding among humans becomes unique and, at the same time, allows the implementation of formalisms to interpret these models appropriately regarding the domain understanding.

\subsection{NLP in Requirements Engineering}
NoRBERT~\cite{HKKT20} is a fine-tuned version of BERT~\cite{DCLT18} for requirements classification. It was trained on the PROMISE NFR dataset \cite{CMLP07}. An approach for clustering natural language requirements is presented in \cite{JNT21}. 
The approach applies clustering to natural language descriptions with the idea of developing a \dsl in mind. 
Apart from requirement classification, machine-learning and \nlp can be used for prioritizing requirements \cite{Perini2013}. These approaches, however, are not specifically tailored for reducing the \emph{initial} modeling efforts necessary for introducing \mdre.

\subsection{Few-Shot Learning for Requirement Classification}
There already exist a few approaches concentrating on few-shot learning in requirements engineering. One of these approaches uses transformer models for a named entity-recognition task~\cite{MCB+22}. Another related approach is a preliminary study on requirement classification using zero-shot learning~\cite{AZF+22}. The PROMISE NFR dataset is used with pre-trained Transformer models such as BERT and RoBERTa. Since it is in a preliminary status, only a reduced part of the dataset is used. In contrast to our approach, the study is only in a preliminary state and zero-shot learning is used for requirement classification instead of few-shot learning. 

Apart from requirement classification, few-shot learning on requirements is also applicable for requirement elicitation as described in~\cite{SXL+20}. In contrast to other approaches in the area of few-shot learning in requirements engineering, the approach does not take already existing requirements as an input, but chat messages from which requirements for new hidden features are extracted. 

\subsection{Research Gap and Goal}
While neural models have been applied to requirements engineering successfully, in particular for requirements classification, there is a research gap concerning the potential of few-shot learning to tailor language models for specific requirements processing tasks such as reformulation, translation, correction, and the like, where training data is scarce, or where access to computational resources needed to fine-tune very large language models such as the GPT-J-6B is limited. The purpose of this work is hence, to develop a set of tailored rules for systematic requirement definition in a given domain by analyzing existing requirements and to determine if and to which extent few-shot learning can support the translation of legacy requirements into the systematic form, thereby aiming to increase the level of formality in requirements.

\section{Example DSL for Structured Requirements}
\label{sec:ReqDSL}
Requirements specify a product under development and thus have a decisive influence on the end product. Missing, incorrect or ambiguously phrased requirements therefore have serious negative effects on the end product and are one of the main causes of additional costs in product development and dissatisfied customers. 

The aim of this section is to design an exemplary requirements \dsl increasing the degree of formality of requirement documents and hence support automated verification and consistency checks. Instead of designing a general requirement language, our approach is to focus on specific domains or projects and to tailor \glspl{dsl} accordingly. Furthermore, we employ few-shot learning capabilities of large neural language models in order to transform existing requirements into the new syntax or to support the formulation of new requirements in an editor. In this work, we focus on the automotive domain, and an existing set of \gls{adas} and \gls{als} requirements from~\cite{BMR+17}.

\dsls provide a means to improve the quality of the requirements' phrasing and to facilitate the implementation of tooling to support agile requirements engineering.
To engineer a \dsl that fulfills the requirements R1-R3 in~\Cref{sec:MDRE}, the \dsl designer must analyze and understand how the requirements are currently phrased and which meanings are implied by certain formulations.

The \dsl is to be used in a natural language domain (\cf R1) and must ensure an intuitive readability for the requirements engineer as well as the modeler who is implementing the requirements at a later stage in the development process, \cf R3. 
The developed \dsl follows an open-world assumption~\cite{DKMR19,DHH+20}, \ie whatever is not restricted by the model is allowed. Currently, the \dsl focuses on isolated requirements written in one sentence. This is because we are mainly interested in concrete syntax and features such as unambiguity. Complex semantic connections and cross-references will not be discussed here.

A manual analysis of the requirements from \cite{BMR+17} reveals a set of ambiguous or unstructured formulations and inconsistencies
which might lead to misinterpretations and hence need to be tackled by the \gls{dsl} approach. The following three ambiguity types are a non-exhaustive selection, which is sufficient as a basis for our few-shot learning experiments.

\textbf{Ambiguity 1: If-Then Constructs}.
The If-Then style is an often reoccurring pattern in requirements documents and an often occurring pattern. It reflects
the idea that upon the occurrence of a trigger event something must happen. In standard English there are numerous variants of how to express this, making it difficult for requirements engineers to stick to a consistent scheme, to search for such requirements, and to analyze them automatically. To tackle this difficulty, the first construct we introduce is the If-Then pattern. It not only creates clarity for the different stakeholders, cf. R3, but also makes further processing in trigger action patterns much easier~\cite{BBB+18}.
Therefore, the \dsl introduces the two keywords \textit{IF:} and \textit{THEN:}. For this purpose, the individual requirements are divided into two parts, a \textit{trigger} part beginning with the keyword \textit{IF:} and an \textit{action} part starting with the keyword \textit{THEN:}, i.e. a parsing rule is given as \texttt{If-Then-Req = 'IF:' trigger 'THEN:' action}. For the sake of simplicity, we assume that the non-terminals \texttt{trigger} and \texttt{action} can be arbitrary strings.

As an example, the requirement \textit{``When no advancing vehicle is recognized anymore, the high beam illumination is restored within 2 seconds.''} is divided into a \textit{trigger} and an \textit{action} part and formulated in the following structure: \textit{``\textbf{IF:} no advancing vehicle is recognized anymore, \textbf{THEN:} high beam illumination is restored within 2 seconds.''}

This way we achieve a partial formalization of the original requirement. The trigger and the action are now clearly separated, and the requirement can be identified as an If-Then requirement easily.

\textbf{Ambiguity 2: Modal Verbs}.
The modal verbs (\texttt{must, can, should, etc.}) are important for the precise interpretation of requirements~\cite{PR21}. Moreover, in safety-critical systems such as vehicles,
there is an important distinction between the available modal verbs. For instance, the modal verb \textit{must} indicates a legal regulation and non-conformity can lead to legal consequences.

Our analysis 
shows however that the model verb is sometimes skipped. In such cases it might not be clear whether the given sentence is a requirement or
a description of an existing system. To enforce the usage of modal verbs we therefore introduce the dedicated keyword \textit{MUST} in the \dsl. In cases of missing modal verbs
the keyword \textit{MUST} needs to be included at the correct position. If a wrong modal verb such as \textit{``can''} or \textit{``could''} is used, it needs to be exchange by \texttt{MUST} thereby preventing the usage of weak words~\cite{PR21}. 

An example is the requirement \textit{``Direction blinking: For the USA and Canada, the daytime running light \textbf{can} be dimmed by 50 percent during direction blinking on the blinking side.''} from the dataset~\cite{BMR+17}. Here, \textit{``can''} has to be replaced by \textit{``MUST''}. 
In the \dsl, the requirement is modeled as \textit{``Direction blinking: For the USA and Canada, the daytime running light \textit{MUST} be dimmed by 50 percent during direction blinking on the blinking side.}

Requirements written without a modal verb should also be supplemented appropriately. 
For example, in the requirement \textit{``The duration of a flashing cycle \textbf{is} 1 second.''}, the \dsl requests a requirement engineer to introduce a \textit{``MUST''} after \textit{``flashing cycle''}. Hence, the model in the \dsl would be: \textit{``The duration of a flashing cycle \textbf{MUST be} 1 second.''}.

\textbf{Ambiguity 3: Expressions.} In requirements, we often need to quantify and compare things. Again, natural language offers many ways to describe comparisons, making it difficult to grasp the information of requirements automatically. For this reason, we introduce a third \gls{dsl} construct for our \gls{dsl}, namely logical formulae. Thereby, we are going to allow both words (to keep the language close to natural formulations) and mathematical expressions in the \dsl. Depending on the use case, the sentences in the \dsl may look different. When using logical signs, these sentences are greatly simplified.

One example is the requirement \textit{``The vehicle's doors are closed automatically when speeding velocity \textbf{is bigger than} 10 km/h''} from~\cite{BMR+17}. 
Modeled in the \dsl integrating logical formulae, the sentence reads as follows: \textit{``The vehicle's doors are closed automatically when speeding velocity $\boldsymbol{>}$ 10 km/h''}. When using the alternative construct for logical formulae with descriptive words offered by the \dsl, the requirement translates to \textit{``The vehicle's doors are closed automatically when speeding velocity is \textbf{GREATER} 10 km/h''}. The difference here is mainly the use of the \dsl. Regarding words, a natural language proximity is desired to increase acceptance by the developer (R1,R2). Further keywords include \textit{``LESS''}, \textit{``EQUAL''}, \textit{``LESS OR EQUAL''}, and \textit{``GREATER OR EQUAL''}.
Again, such \gls{dsl} constructs standardize the 
way how comparisons are formulated in a requirement. This facilitates automated consistency analysis since the operators and their semantics are clearly defined and 
the variables and constants can be extracted relatively easy from the sentence, e.g. \textit{velocity} and \textit{10 km/h}. Now,
if the same variable is used in another requirement, we can 1) link these requirements
as treating the same context, e.g. to enable semantic requirement search and 2) are able to check whether the two requirements are consistent.

The three constructs introduced above can and should be combined when appropriate. For instance, the last example needs to be translated to \textit{``IF: speeding velocity is \textbf{GREATER} 10 km/h, THEN: the vehicle's doors MUST be closed automatically.''}

Now this requirement has almost no degrees of freedom in terms of formulation, and it is easy to extract the trigger variable (speeding velocity), the subject of the action (the vehicle's doors), and the desired state (closed automatically).

The challenge we would like to solve in this paper is how to translate large numbers of legacy requirements into the \gls{dsl}, how to repair requirements showing syntactical errors, and how to support a requirement engineer writing requirements in the \gls{dsl} syntax.

\section{Translating Unstructured Requirements to DSL}
\label{sec:Translation}
\subsection{Overview}
To automatically translate legacy natural language requirements into the \gls{dsl} defined in \Cref{sec:ReqDSL}, fixing bad formulations and enforcing guidelines usage and regulatory compliance, we utilize transformer-based language models. 
Since we need to avoid data and resource intensive finetuning (as the necessary amounts of data might lack in project or company-specific design and the required hardware resources might not be accessible/too expensive), we will make use of the few-shot capability, which has been shown to yield good results with only a few training examples. Hence, we rule out BERT and RoBERTa as models for our translation task and stick to the GPT family. 

The number of training examples for few-shot learning a new task is usually constrained by the context window, typically allowing 10-100 examples~\cite{BMR+20}. Sometimes the number of examples is further constrained by the available computational resources. To exploit the available training data as far as possible, we propose a cascaded translation process, where we provide a dedicated few-shot model for each translation task, i.e. trigger-action, modal verbs, and expressions. This reduces the training complexity and yields more focused models as the neural network does not have to decide, which kind of translation(s) to perform but only has to perform the chosen translation it is specialized for. 

In case, a single translation needs to be applied, the translation step needs to be preceded either by a manual or an automated classification of the
input and the corresponding choice of a translation model. Otherwise a sequential
application of multiple models is necessary to incorporate all \gls{dsl} constructs.

\subsection{Requirement Translation}
\label{sec:RequirementsTranslation}
For each translation step, a set of few-shot examples, also referred to as the support set, for the respective requirements category is selected and given to the language model as context. Our hypothesis is
that a large capacity language model pretrained on a sufficiently large training set can be taught to solve a specific task such as a reformulation of a given requirement into a systematic form with a very small
support set. In contrast to fine-tuning approaches, where a pretrained model is further trained using a downstream task-specific training set, in few-shot learning the weights of the neural network are not adapted. The few-shot examples consist of input/output pairs and are input into the network as a demonstration for the task to be solved, followed by the actual query. The solution to the query is then generated as the model's output based on the support examples from the context. 

The translation model can be implemented using any large enough pretrained language model supporting few-shot learning. Models of higher capacity can be expected to perform better in few-shot learned downstream tasks. Based on promising preliminary results compared to other models, we decided to concentrate mainly on GPT-J-6B. GPT Neo \cite{gpt-neo}, another model from the
GPT family, showed a lot of variance in its text generation
producing a lot of "noisy", unexpected outputs often changing or extending the semantics of the source requirement
without an obvious reason in our preliminary experiments. 

\section{Evaluation}
\label{sec:Evaluation}
\subsection{Research Questions and Metrics}
With the experiments conducted for the translation of requirements from unstructured text to \gls{dsl} our aim was to find answers to the following research questions:

\begin{enumerate}
    \item[\textbf{RQ1:}] Can state-of-the-art language models
be employed to translate natural text requirements to systematic formulations based on few-shot learning?

    \item[\textbf{RQ2:}] How many few-shot learning examples are required to train a translation rule?
\end{enumerate}

To answer RQ1, we applied few-shot learning separately for If-Then requirements, modal verb insertion, and expressions (\cf~\Cref{sec:ReqDSL}) (recall that for training each rule we use a separate instance of the language model). 
As introduced in~\Cref{sec:ReqDSL} the trained constructs are:
\begin{enumerate}
    \item[a)] ``\textit{'IF:' trigger 'THEN:' action}'' syntax for trigger action requirements, 
    \item[b)] modal verbs enforcement, reframing sentences to contain the ``\textit{MUST}'' keyword, and 
    \item[c)] the usage of ``\textit{LESS / GREATER / EQUAL / LESS EQUAL / GREATER EQUAL}'' operators in constraints.
\end{enumerate}

Input for the few-shot learning were pairs containing the unstructured input and the desired DSL formulation. 
To evaluate the ``trained'' language model, it had to transform an unstructured requirement that the model was not given as example into a requirement in the \dsl. 
We then assessed the result of this transformation.
Doing so, requires a significant evaluation metric.
Metrics such as BLEU \cite{PRW+02} typically used to evaluate results
of machine translation are not helpful, due to the many different possibilities to express a requirement correctly. Furthermore, semantic correctness does not imply compliance with the expected
syntactical structure we would like to enforce. For this reason, we propose
a custom evaluation scale with six possible quality classes
to assess the correctness of the translated requirements:

\begin{enumerate*}
    \item[\textbf{Class 1}:] The translation is both syntactically and semantically correct and fulfills the required formulation rule. No changes required.
    \item[\textbf{Class 2}:] The translation is semantically correct, but contains one or two syntactical inaccuracies to fully implement the desired rule.
    \item[\textbf{Class 3}:] Syntactically correct but fails to fully cover the semantics of the source requirement (e.g. by missing a quantifier or a marginal constraint).
    \item[\textbf{Class 4}:] The translation contains one or two syntactical inaccuracies to fully implement the desired rule and the semantics is not fully covered, i.e. a combination of 2 and 3.
    \item[\textbf{Class 5}:] The translation has grave syntactical errors or does not implement the desired rule. An identity mapping would result in this label, as well (unless the input already implements the desired rule).
    \item[\textbf{Class 6}:] The translation is semantically wrong.
\end{enumerate*}

A flaw of this scale is that it is not ordinal, e.g. we cannot say that class 2 is better than class 3. However, based on the experience we gathered with it in this work, in most cases a smaller number indicates a more satisfying result.

To answer RQ2 we conducted our evaluation with three
differently sized few-shot support sets per translation rule (If-Then, modal verbs, and constraints) consisting of one, four, and six examples each. In case of one-shot learning, \ie if only one example is presented in training, for the conversion of constraints containing (in)equalities, the result depends on the keyword used in the example. For this reason, we tried two different one-shot trainings, \ie for introducing ``\textit{EQUAL}'', and ``\textit{LESS OR EQUAL}''. The requirements used for testing were \textbf{not} present in the support sets.

\subsection{Concrete Training Examples}
In this section we provide excerpts of the training and prediction experiments to give the reader an intuition
on how the GPT-J-6B model dealt with the translation, which constructs were particularly challenging, and how the number of examples affected the quality of the translation. The full support sets and the complete translation experiment results can be found in the appendix in \cref{tab:supset:logic:1:equal}-\cref{tab:eval:modal:6}.

\textbf{If-Then, 1 training example:} To one-shot train the translation of a requirement to the desired If-Then syntax we use the following input and output pair:

\noindent \textit{``Input: If a defective illuminant is detected, the information about the defective illuminant is transmitted to the instrument cluster.\\
DSL: IF: defective illuminant is detected, THEN: information about the defective illuminant is transmitted to the instrument cluster.''}

As we can see in the example, the two keywords \textit{``IF:''} and \textit{``THEN:''} are included in the target sentence. Apart from that the sentence remains almost unchanged, making it relatively easy for the model
to learn the rule.

Having seen this example, the model is already able to apply the rule perfectly to some examples of the test set, yielding a class 1 rating according to the scheme given above. For instance, the requirement

\noindent \textit{``If tip-blinking was activated shortly before deactivation of the hazard warning, this is not considered during the deactivation of the hazard warning.''} 

\noindent is correctly translated to

\noindent \textit{``IF: tip-blinking was activated shortly before deactivation of the hazard warning, THEN: this is not considered during the deactivation of the hazard warning
''}.

However, some other examples are translated incompletely or wrong, e.g. the requirement 

\noindent \textit{``With activated darkness switch (only armored vehicles) the cornering light is not activated.''}

\noindent is translated to 

\noindent \textit{``IF: darkness switch is activated, THEN: cornering light is not activated.
''} 

While the keywords are included at the correct position, the model drops the information in parentheses
stating that the requirement only applies to armored vehicles. For this reason, we consider this translation as syntactically correct, but semantically incomplete, resulting in a class 3 rating.

At this point the model has not yet fully learnt to implement the If-Then syntax if the sentence does not start with an \textit{``if''}. For instance, the example 

\noindent \textit{``When moving the pitman arm in position turn left the vehicle flashes all left direction indicators (front left, exterior mirror left, rear left) synchronically with pulse ratio bright to dark 1:1.''} 

\noindent remains
unchanged by the one-shot model. This yields a class 5 rating, as the semantics is correct, but the desired syntax is not implemented. Interestingly, another requirement starting with \textit{``when''} (\textit{``When hazard warning is deactivated and the pit arm is in position direction blinking left or direction blinking right, the direction blinking cycle should be released.
''}) results in the output 

\noindent \textit{``WHEN: hazard warning is deactivated, THEN: direction blinking cycle should be released
''}. 

Here the model introduces a new keyword (\textit{``WHEN:''}), which did not happen in the previous example. Furthermore the second condition concerning the pit arm position has been dropped, resulting in a class 6 rating due to both syntactical and semantical errors.

\textbf{If-Then, 4 training examples:} Aiming to improve the translation quality, we introduce three additional training example pairs:

\noindent \textit{``Input: If no advancing vehicle is recognized any more, the high beam illumination is restored within 2 seconds. \\
DSL: IF: no advancing vehicle is recognized any more, THEN: high beam illumination is restored within 2 seconds.\\
\#\#\#\\
Input: If the light rotary switch is in position "auto", the adaptive high beam headlights are activated by moving the pitman arm to the back. \\
DSL: IF: the light rotary switch is in position "auto" and the pitman arm is moved back, THEN: the adaptive high beam headlights are activated.\\
\#\#\#\\
Input: (a) When the driver enables the cruise control (by pulling the cruise control lever or by pressing the cruise control lever up or down), the vehicle maintains the set speed if possible. \\
DSL: IF: driver enables the cruise control by pulling the cruise control lever or by pressing the cruise control lever up or down, THEN: the vehicle maintains the set speed if possible.
''}

The three additional examples lead to an improvement in translation quality, for instance, instead of generating a new keyword \textit{``WHEN:''}, the model now knows due to the last training example that it has to replace \textit{``When''} by \textit{``IF:''}. So now, the hazard warning requirement is translated correctly to 

\noindent \textit{``IF: hazard warning is deactivated and the pit arm is in position direction blinking left or direction blinking right, THEN: the direction blinking cycle should be released.''} and yields a class 1 rating.

\textbf{If-Then, 6 training examples:} To further improve the translation quality, the following examples have been added:

\noindent \textit{``
Input: If the driver pushes down the cruise control lever with cruise control activated up to the first resistance level, the speed set point of the cruise control is reduced by N. \\
DSL: IF: driver pushes down the cruise control lever with cruise control activated up to the first resistance level, THEN: the speed set point of the cruise control is reduced by N. \\
\#\#\#\\
Input: By pushing the brake or the hand brake, the cruise control is deactivated until it is activated again. \\
DSL: IF: the brake or the hand brake is pushed, THEN: the cruise control is deactivated until it is activated again.''}

In this variant the translation of the armored vehicle requirement is improved to a class 2 rating yielding
the text \textit{``IF: with activated darkness switch (only armored vehicles), THEN: the cornering light is not activated.
''} Here the model manages to keep the information on armored vehicles in parentheses, but chooses a strange formulation in 
the if-part, i.e. small syntactic fixes are required. Hence, the model improves from a class 3 to a class 2 grade.

Requirements of the form \textit{``Context: requirement text.''} were particularly problematic for the model. The context was often interpreted as a condition. For instance, the requirement \textit{``Distance Warning: The vehicle warns the driver visually and/or acoustically if the vehicle is closer to the car ahead than allowed by the safety distance.''} is translated to \textit{``IF: distance warning is activated, THEN: the vehicle warns the driver visually and/or acoustically''}. This problem persists for all sizes of support sets
we have tried, yielding a class 6 grading. Probably, the problem also persists for other deviating sentence structures.

\textbf{Modal verb, 1 training examples:}
The one-shot support set for the insertion of modal verbs is given as: 

\noindent \textit{``Input: The maximum deviation of the pulse ratio should be below the cognitive threshold of a human observer.\\
DSL: The maximum deviation of the pulse ratio MUST be below the cognitive threshold of a human observer.''}

Given this single example the model is able to insert the \textit{``MUST''} keyword into simple sentences. For instance, the descriptive sentence \textit{``The frame rate of the camera is 60 Hz.''} is translated to a
real requirement \textit{``The frame rate of the camera MUST be 60 Hz.''}, thereby yielding a class 1 grading. However, at this point it fails to insert the \textit{``MUST''} keyword in most other test examples.

\textbf{Modal verb, 4 training examples:''} For the 4-shot support set we use the following additional training example pairs:

\noindent \textit{``Input: Direction blinking: For USA and CANADA, the daytime running light shall be dimmed by 50\% during direction blinking on the blinking side.\\
DSL: Direction blinking: For USA and CANADA, the daytime running light MUST be dimmed by 50\% during direction blinking on the blinking side.\\
\#\#\#\\
Input: The adaptation of the pulse ratio has to occur at the latest after two complete flashing cycles. \\
DSL: The adaptation of the pulse ratio MUST occur at the latest after two complete flashing cycles. \\
\#\#\#\\
Input: The duration of a flashing cycle is 1 second.\\
DSL: The duration of a flashing cycle MUST be 1 second.''}

This helps the model to translate more inputs correctly, e.g. while the requirement \textit{``A flashing cycle (bright to dark) has to be always completed, before a new flashing cycle can occur''} remained unchanged by the model in the one-shot case, it is now translated to \textit{``A flashing cycle (bright to dark) MUST be always completed, before a new flashing cycle can occur.''} This yields a class 1 grade.

Similarly, the example \textit{``Also after 1000 flashing cycles the cumulated deviation will not exceed 0.05s.''} is now translated to textit{``Also after 1000 flashing cycles the cumulated deviation MUST NOT exceed 0.05s.''}, again yielding class 1. 

\textbf{Modal verb, 6 training examples:}
Adding the following additional two training examples to the support set further improves the translation quality:

\noindent \textit{``Input: A flashing cycle (bright to dark) will always be completed, before a new flashing cycle can occur. \\
DSL: A flashing cycle (bright to dark) MUST always be completed, before a new flashing cycle can occur.\\
\#\#\#\\
Translate input to DSL
Input: With subvoltage the ambient light is not available. \\
DSL: With subvoltage the ambient light MUST not be available.''}

The two inputs \textit{``With subvoltage the ambient light is not available.''} and \textit{``Low beam illuminant shall be LED.''} are now correctly translated to \textit{``With subvoltage the ambient light MUST not be available.''} and \textit{``Low beam illuminant MUST be LED.''}.

Some inputs are still not translated correctly, e.g. \textit{``The functions of the system are classified as safety relevant in with respect to ISO 26262.''} But also the seemingly simpler requirement formulation \textit{``The vehicle does not exceed a set speed.''}.

\textbf{Expressions with keywords, 1 training examples:}
In the next experiment, we aim to detect expressions, in particular comparisons in requirements and
insert keywords such as \textit{``EQUALS''} or \textit{``GREATER''} into the sentences.
The one-shot support set is given as:

\noindent \textit{``Input: The maximum deviation of the pulse ratio should be below the cognitive threshold of a human observer.\\
DSL: The maximum deviation of the pulse ratio should be LESS OR EQUAL cognitive threshold of human observer.''}

The problem here is the relatively large number of keywords. The model cannot infer all the keywords from a single example. Nevertheless, it tries to apply the \textit{``LESS OR EQUAL''} keyword 
to some test inputs at the correct place. For instance, the requirement 

\noindent \textit{``The luminous intensity of the daytime running light must be lower than 400cd.''} 
\noindent is transformed to

\noindent \textit{``The luminous intensity of the daytime running light must be LESS OR EQUAL 400cd.''}. 

\noindent The result has a semantic flaw as it allows
an equality which is not the case for the original, hence resulting in a class 3 rating.

Interestingly, the model transforms \textit{``The maximum curb weight of the vehicle must be no more than 3.5t.''} to \textit{``The maximum curb weight of the vehicle must be LESS THAN 3.5t.''}, thereby introducing a keyword it has not seen in training. In contrary to the previous requirement, 
equality is allowed here in the original but is excluded by the mode, again resulting in a lass 3 rating.

We tried a one shot variant with an exampling using the \textit{``EQUAL''} keyword instead of \textit{``LESS OR EQUAL''}. As expected, the model then only used this keyword and hence, performed better
where \textit{``EQUAL''} was required and worse otherwise.

\textbf{Experessions with keywords, 4 training examples:} In the next experiment we used the four following examples as support set:

\noindent \textit{``Input: The duration of a flashing cycle is 1 second
DSL: The Duration of a flashing cycle is EQUAL 1 second.\\
\#\#\#\\
Input: The maximum deviation of the pulse ratio should be below the cognitive threshold of a human observer.\\
DSL: The maximum deviation of the pulse ratio should be LESS OR EQUAL cognitive threshold of human observer.
\\
\#\#\#\\
Input: The minimal number of seatbelts used has to be 1.\\
DSL: The number of seatbelts used has to be GREATER OR EQUAL 1.
\\
\#\#\#\\
Input: The vehicles doors are closed automaticly when speeding velocity is bigger than 10km/h.\\
DSL: The vehicles doors are closed automaticly when speeding velocity is GREATER 10km/h.''}

Surprisingly, the model was biased towards inserting the \textit{``LESS THAN''} keyword, which was not present in the support set at all. The overall improvements were minor. The luminous intensity requirement
was now transformed correctly to \textit{``The luminous intensity of the daytime running light must be LESS THAN 400cd.''}, yielding a class 1 rating as the only requirement.

\textbf{Experessions with keywords, 6 training examples:}
For the last experiment we added the following two training examples:
\noindent \textit{``Input: Also after 1000 flashing cycles the cumulated deviation must not exceed 0.05s.
DSL: After 1000 flashing cycles the cumulated Deviation must be GREATER OR EQUAL 0.05s.
\\
\#\#\#\\
Input: The cruising speed has to be set at a speed which exceeds 10km/h. 
DSL:  The cruising speed has to be GREATER 10km/h.''}

In this experiment additionally the maximum curb weight requirement was transformed perfectly to \textit{`` The maximum curb weight of the vehicle must be LESS OR EQUAL 3.5t.''}, now including the equality part.

The experiment shows a main drawback inherent to models such as GPT. The models use statistically
learned sequences without understanding the semantics. For instance, the model often confuses less-than and greater-than inequalities, which are very similar from the syntactical point of view, but have a completely different semantics. For instance, the model delivers translations such as \textit{``The minimum distance to a vehicle in front has to be LESS OR EQUAL 5m.''}. While it is obvious to a human reader that \textit{``GREATER''} should be used here, a neural language model has no logical interpretation.

Another problem which becomes evident in this evaluation is that the language model has difficulties
reformulating sentences when a pure insertion of a keyword is insufficient. For instance,
the example \textit{``The vehicle's horn must not be louder than 50dB.''} is transformed to \textit{``The vehicle's horn must be LESS OR EQUAL 50dB.''}. While the keyword used is correct, the sentence has a semantic flaw: it is not the horn that has to be less or equal 50dB but its intensity, i.e. the perfect translation would be \textit{```The intensity of the vehicle's horn must be LESS OR EQUAL 50dB.''}. The model struggles to insert the physical
quantity (intensity) here.

\textbf{Expression Extraction}
Additionally, we performed experiments aiming to extract mathematical formulas only containing the key operators and variable names. For instance, the input \textit{``The braking distance can not be longer than 300m.''} yields the output \textit{``braking distance <= 300m''}. Such an extraction can be useful to 
implement tracing and consistency checking of multiple requirements. For instance, if the variable braking distance is found in multiple requirements and featuring different bounds, this can be found automatically
and shown to a human expert to check whether the requirements are indeed contradictory. Further successful
extractions we generated using the model are \textit{``horn loudness <= 50dB''}, \textit{``low beam illuminant = LED''}, and \textit{``distance to vehicle in front >= 5m''}. Most others showed slight to medium
inacuracies, e.g. \textit{``maximum velocity <= 260km/h''} or \textit{``blinking lights = 3s''}.

\subsection{Evaluation Summary}

An overview of all experiment results is summarized in~\Cref{tab:translation}.
As expected, in each of the three experiments, the translation quality improved with larger support sets. It is fascinating however, how steep the learning curve is. 
It suggests that few-shot learning can deal with \gls{nlp} tasks in requirements engineering even when only small training sets are available.

\begin{table*}
\caption{\label{tab:translation}Evaluation results for the translation experiments from
natural language requirements to domain-specific syntax.}
\centering
\resizebox{0.70\textwidth}{!}{%
\begin{tabular}{ |p{3cm}||p{1,5cm}|p{1,5cm}|p{1,5cm}|p{1,5cm}|p{1,5cm}|p{1,5cm}|p{1,5cm}|}
 \hline
 \# of Training Set & Class 1 & Class 2 & Class 3 & Class 4 & Class 5 & Class 6 & total \\
 \hline
 \multicolumn{8}{|c|}{Translation results for If-Then structure using GPT-J6B} \\
 \hline
  \hline

1 & 2 & 0 & 1 & 0 & 6 & 2 & 11 \\
 \hline

4 & 5 & 2 & 0 & 1 & 1 & 2 & 11 \\ 
 \hline

6 & 6 & 3 & 0 & 0 & 0 & 2 & 11 \\
 \hline

 \hline
 \multicolumn{8}{|c|}{Translation results for modal verbs structure using GPT-J6B} \\
 \hline
  \hline

1 & 3 & 0 & 0 & 0 & 5 & 0 & 8 \\
 \hline

4 & 5 & 0 & 0 & 0 & 3 & 0 & 8 \\ 
 \hline

6 & 6 & 0 & 0 & 0 & 2 & 0 & 8 \\
 \hline

 \hline
 \multicolumn{8}{|c|}{Translation results for propositional logic Structure using GPT-J6B} \\
 \hline
  \hline

1 (trained on keyword: equal) & 2 & 0 & 1 & 0 & 4 & 1 & 8 \\
 \hline
1 (trained on keyword: less or equal) & 0 & 0 & 4 & 0 & 4 & 0 & 8 \\
 \hline

4 & 1 & 0 & 1 & 0 & 3 & 3 & 8 \\ 
 \hline

6 & 2 & 0 & 3 & 0 & 2 & 1 & 8 \\
 \hline
8 & 3 & 0 & 0 & 0 & 3 & 2 & 8 \\
 \hline
\end{tabular}}
\end{table*}

In the first experiment (If-Then translation), given a support set of six training examples, six out of eleven examples achieve a class 1 rating, \ie the model did a perfect translation. Further, the results with only four training pairs in the support set were still good. However, two requirements were transformed incorrectly. An example for a class 1 translation is: \textit{``Emergency Brake Assist: The vehicle decelerates in critical situations to a full standstill.''} is transformed to \textit{``IF: emergency brake assist is activated, THEN: the vehicle decelerates in critical situations to a full standstill.''} Here the malformed sentence is restructured with the verb ``is activated'', separating trigger and action. Furthermore, the model learns to replace words such as ``if'', ``once'', ``when'' by the keyword ``IF:'' and to insert ``THEN:'' at the correct position. The two requirements which were not transformed correctly even for six training examples both contained two conditions and were hence more difficult to translate. Apparently, the network has difficulties processing inputs consisting of several sentences, e.g. multiple trigger-action clauses. However, subdividing such requirements into separate inputs seems to solve the problem.

The second experiment (modal verbs translation) was probably the easiest, which is not surprising. The language model only needs to add or replace the right verb with ``MUST''. For instance ``The frame rate of the camera is 60 Hz.'' was successfully translated to ``The frame rate of the camera MUST be 60 Hz''. The model did this well for all but two requirements given six training examples. One of the two outliers contained a negation, which was not explicitly trained.

For the third experiment, dealing with constraints (introducing (in-)equalities), to explore the capability of our model to identify and generate constraint sentences,
we crafted an additional dataset consisting of six few-shot training and six test examples featuring equality and inequality clauses. Due to more variance in the desired translations (five keywords to choose from) and the difficulty encountered by the model to transform adjectives to subjects (e.g. ``the horn has to be louder than'' needs to become ``the horn intensity has to be GREATER than''), we added three more training examples focusing on this issue. It did however not lead to the desired results, the model did not learn to transform the sentence accordingly. Sentences for which no such adjective-to-subject transformation was necessary 
had a better success rate in the test set. For instance, ``The maximum curb weight of the vehicle must be no more than 3.5t.'' was successfully translated to ``The maximum curb weight of the vehicle must be LESS OR EQUAL 3.5t.'' Some translations featured a wrong operator, but were otherwise syntactically correct. In applications such as smart editors, this could still be a helpful way to sensitize the requirement engineers for the correct syntax.

We conducted a variant of the third experiment, where a) the natural language operators were replaced by mathematical ones, i.e. $<, \leq, >, \geq, =$, and  b) only the variables/constants of the (in-)equality were extracted in the support set, i.e. without keeping the rest of the sentence. The model managed to perform these transformations surprisingly well (except for one example where the model put a ``$\leq$'' instead of ``$=$'' for the maximum velocity), e.g. from ``The vehicles horn must not be louder than 50dB'' the model extracted ``horn loudness $\leq$ 50dB''. Such extractions can be used in tracing, e.g. to identify  assignments or conditions in code which are inconsistent with the underlying requirements. Variables and their values or bounds can be  extracted by a language model and looked up in code automatically.

\section{Threats to Validity}\label{sec:threats}
\paragraph{Construction Validity}
An objective metric of requirement classification and translation is not feasible per se and depends on the context, perspective of the reader, and other factors. For our evaluation we let five software engineers label the translations independently and took the majority vote after a discussion to mitigate the risk of bias. 

\paragraph{Internal Validity}
Our evaluation dataset might have a too small variance:
The legacy examples stem from a single project and are mostly well written, there are no completely ill-formed examples. This might render the translation process too easy for the language model. 

\paragraph{External Validity}
Generalizability is a major concern in requirements engineering as findings and algorithms need to be transferred to unseen projects and domains. The methodology presented in this paper is designed for domain-specific adaptations, \ie the results are only valid for the datasets at hand.
For the translation task, we cannot rule out that different or more complex syntactic elements required in other domains cannot be realized using few-shot learning  or require larger support sets. Generalizability of the concrete few-shot models derived in this paper across different domains is probably not feasible due to differences in domain-specific wordings and formulation approaches, but also not pursued in this work since each domain should use tailored support sets. 

\section{Conclusion}
\label{sec:Conclusion}
In this paper, we have shown how neural language models such as representatives of the GPT family can support requirements engineering without the need for resource and data intensive fine-tuning.
Our most important result is that few-shot learning of language models can be applied to translate legacy requirements into 
a given structured \gls{dsl} form automatically. However, language models available today still require human supervision. Since context space available in
a language model for few-shot learning is limited to a very small number of examples, we suggest to manually or automatically classify a requirement before translation to choose the few-shot examples to be used for a given translation task. 

We draw the conclusion that few-shot learning is a powerful tool for the processing
of natural language requirements which does not require costly training from scratch and expensive
hardware and is hence accessible even for small companies with limited resources or when training data is scarce. It has the potential to accelerate the formalization of legacy requirements and improve the resulting requirements quality and syntactical consistency. However, as of now, supervision cannot be eliminated, binding human resources but limiting the task to correction. 

Instead of having to rely on the generalizability of
a single language model for all projects, the few-shot learning approach can be 
employed to tailor language models to new projects and application domains quickly. In very specific areas, singular fine-tunings to the domain might be necessary, e.g. we realized in our experiments that in some cases the language models had difficulties processing automotive-related language terms correctly. 

We believe that the presented technology can be used for a variety of tasks in model-driven requirements
engineering including but not limited to the translation of legacy requirements, autocompletion for smart editors, requirement look-up and information extraction. While there is still room for improvement, it is reasonable to expect that the capacity of language models will steadily grow larger in the next years providing a continuously increasing few-shot learning performance.

  \bibliographystyle{ACM-Reference-Format}
  \bibliography{tex/bib/sselit,tex/bib/idr,tex/bib/ek,tex/bib/lw,tex/bib/vb}

\newpage
\onecolumn
\appendix
\section{Appendix}
\label{sec:Appendix}
\setcounter{table}{0}
\renewcommand\thetable{\Alph{section}.\arabic{table}}

\begin{table}[h]
\centering
\caption{Support Set: 1 Example Logical Formulae (EQUAL).}
\begin{tabularx}{0.95\textwidth}{|X|X|}
\hline
\textbf{Original}         & \textbf{Translation} \\\hline
The duration of a flashing cycle is 1 second.                 & The duration of a flashing cycle is EQUAL 1 second. \\ \hline
\end{tabularx}
\label{tab:supset:logic:1:equal}
\end{table}

\begin{table}[h]
\centering
\caption{Support Set: 1 Example Logical Formulae (LESS-OR-EQUAL).}
\begin{tabularx}{0.95\textwidth}{|X|X|}
\hline
\textbf{Original}         & \textbf{Translation} \\\hline
The maximum deviation of the pulse ratio should be below the cognitive threshold of a human observer                & The maximum deviation of the pulse ratio should be LESS OR EQUAL cognitive threshold of human observer. \\ \hline
\end{tabularx}
\label{tab:supset:logic:1:lessorequal}
\end{table}

\begin{table}[h]
\centering
\caption{Support Set: 4 Examples Logical Formulae.}
\begin{tabularx}{0.95\textwidth}{|X|X|}
\hline
\textbf{Original}         & \textbf{Translation} \\\hline
The duration of a flashing cycle is 1 second.               & The duration of a flashing cycle is EQUAL 1 second. \\ \hline
The maximum deviation of the pulse ratio should be below the cognitive threshold of a human observer. &
The maximum deviation of the pulse ratio should be EQUAL cognitive threshold of human observer.\\\hline
The minimal number of seatbelts used has to be 1. & The number of seatbelts used has to be GREATER OR EQUAL 1.\\\hline
The vehicles doors are closed automaticly when speeding velocity is bigger than 10km/h. & The vehicles doors are closed automaticly when speeding velocity is GREATER 10km/h.\\ \hline
\end{tabularx}
\label{tab:supset:logic:4}
\end{table}

\begin{table}[h]
\centering
\caption{Support Set: 6 Examples Logical Formulae.}
\begin{tabularx}{0.95\textwidth}{|X|X|}
\hline
\textbf{Original}         & \textbf{Translation} \\\hline
The duration of a flashing cycle is 1 second.              & The duration of a flashing cycle is EQUAL 1 second.
 \\ \hline
Also after 1000 flashing cycles the cumulated deviation must not exceed 0.05s. & After 1000 flashing cycles the cumulated Deviation must be GREATER OR EQUAL 0.05s. \\\hline
The cruising speed has to be set at a speed which exceeds 10km/h. & The cruising speed has to be GREATER 10km/h.
\\\hline
The maximum deviation of the pulse ratio should be below the cognitive threshold of a human observer. & 
The maximum deviation of the pulse ratio should be EQUAL cognitive threshold of human observer. \\\hline
The minimal number of seatbelts used has to be 1. & The number of seatbelts used has to be GREATER OR EQUAL 1. \\\hline
The vehicles doors are closed automaticly when speeding velocity is bigger than 10km/h. & The vehicles doors are closed automaticly when speeding velocity is GREATER 10km/h.
\\\hline

\end{tabularx}
\label{tab:supset:logic:6}
\end{table}

\begin{table}[h]
\centering
\caption{Support Set: 8 Examples Logical Formulae.}
\begin{tabularx}{0.95\textwidth}{|X|X|}
\hline
\textbf{Original}         & \textbf{Translation} \\\hline
The duration of a flashing cycle is 1 second.              & The duration of a flashing cycle is EQUAL 1 second.
 \\ \hline
Also after 1000 flashing cycles the cumulated deviation must not exceed 0.05s. & After 1000 flashing cycles the cumulated Deviation must be GREATER OR EQUAL 0.05s. \\\hline
The cruising speed has to be set at a speed which exceeds 10km/h. & The cruising speed has to be GREATER 10km/h.
\\\hline
The maximum deviation of the pulse ratio should be below the cognitive threshold of a human observer. & 
The maximum deviation of the pulse ratio should be EQUAL cognitive threshold of human observer. \\\hline
The minimal number of seatbelts used has to be 1. & The number of seatbelts used has to be GREATER OR EQUAL 1. \\\hline
The vehicles doors are closed automaticly when speeding velocity is bigger than 10km/h. & The vehicles doors are closed automaticly when speeding velocity is GREATER 10km/h.
\\\hline
The interior material must be Velour.  & The interior material is EQUAL Velour. \\\hline
Flashing light cycle does not exceed 3 flashes. & Flashing light cycle is LESS or EQUAL 3 flashes.
\\\hline
\end{tabularx}
\label{tab:supset:logic:8}
\end{table}

\begin{table}[h]
\centering
\caption{Evaluation Results: 1 Example Logical Formulae (EQUAL).}
\begin{tabularx}{0.95\textwidth}{|X|X|l|}
\hline
\textbf{Original}         & \textbf{Translation} & \textbf{Class} \\\hline
The luminous intensity of the daytime running light must be lower than 400cd. & The luminous intensity of the daytime running light must be EQUAL 400cd. & 6 \\\hline
The range of the vehicle shall be at least 450km under WLTP conditions & The range of the vehicle shall be at least 450km under WLTP conditions. & 5 \\\hline
If the speed limiter is activated, the vehicle speed shall not exceed the set speed.&If the speed limiter is activated, the vehicle speed shall not exceed the set speed.& 5 \\\hline
The maximum curb weight of the vehicle must be no more than 3.5t & The maximum curb weight of the vehicle must be EQUAL 3.5t. & 3\\\hline
The breaking distance can not be longer than 300m.  & The breaking distance can not be longer than 300m. & 5 \\\hline
The vehicles horn must not be louder than 50dz. &The vehicles horn must not be louder than 50dz. &5\\\hline
The vehicles maximum velocity is 260km/h.  &The vehicles maximum velocity is EQUAL 260km/h.&1\\\hline
The minimun distance to a vehicle in front has to be 5m. &The minimun distance to a vehicle in front has to be EQUAL 5m. &1\\\hline
\end{tabularx}
\label{tab:eval:logic:1:equal}
\end{table}

\begin{table}[h]
\centering
\caption{Evaluation Results: 1 Example Logical Formulae (LESS-OR-EQUAL).}
\begin{tabularx}{0.95\textwidth}{|X|X|l|}
\hline
\textbf{Original}         & \textbf{Translation} & \textbf{Class} \\\hline
The luminous intensity of the daytime running light must be lower than 400cd. & The luminous intensity of the daytime running light must be LESS OR EQUAL 400cd.  & 3 \\\hline
The range of the vehicle shall be at least 450km under WLTP conditions & The range of the vehicle shall be at least 450km under WLTP conditions.  & 5 \\\hline
If the speed limiter is activated, the vehicle speed shall not exceed the set speed.&If the speed limiter is activated, the vehicle speed shall not exceed the set speed.& 5 \\\hline
The maximum curb weight of the vehicle must be no more than 3.5t & The maximum curb weight of the vehicle must be LESS THAN 3.5t. & 3\\\hline
The breaking distance can not be longer than 300m.  & The breaking distance can not be longer than 300m. & 5 \\\hline
The vehicles horn must not be louder than 50dz. &The vehicles horn must not be louder than 50dz. &5\\\hline
The vehicles maximum velocity is 260km/h.  &The vehicles maximum velocity is LESS OR EQUAL 260km/h. &3\\\hline
The minimun distance to a vehicle in front has to be 5m. & The minimum distance to a vehicle in front has to be LESS OR EQUAL 5m.  &3\\\hline
\end{tabularx}
\label{tab:eval:logic:1:lessorequal}
\end{table}

\begin{table}[h]
\centering
\caption{Evaluation Results: 4 Examples Logical Formulae.}
\begin{tabularx}{0.95\textwidth}{|X|X|l|}
\hline
\textbf{Original}         & \textbf{Translation} & \textbf{Class} \\\hline
The luminous intensity of the daytime running light must be lower than 400cd. & The luminous intensity of the daytime running light must be LESS THAN 400cd.  & 1 \\\hline
The range of the vehicle shall be at least 450km under WLTP conditions & The range of the vehicle shall be at least 450km under WLTP conditions.  & 5 \\\hline
If the speed limiter is activated, the vehicle speed shall not exceed the set speed.& If the speed limiter is activated, the vehicle speed shall not exceed the set speed.& 5 \\\hline
The maximum curb weight of the vehicle must be no more than 3.5t &  The maximum curb weight of the vehicle must be LESS THAN 3.5t. & 3\\\hline
The breaking distance can not be longer than 300m.  &  The breaking distance can not be LESS THAN 300m. & 6 \\\hline
The vehicles horn must not be louder than 50dz. & The vehicles horn must not be LOUDER THAN 50dz. &5\\\hline
The vehicles maximum velocity is 260km/h.  & The vehicles maximum velocity is LESS THAN 260km/h. &6\\\hline
The minimun distance to a vehicle in front has to be 5m. &  The minimun distance to a vehicle in front has to be LESS THAN 5m.  &6\\\hline
\end{tabularx}
\label{tab:eval:logic:4}
\end{table}

\begin{table}[h]
\centering
\caption{Evaluation Results: 6 Examples Logical Formulae.}
\begin{tabularx}{0.95\textwidth}{|X|X|l|}
\hline
\textbf{Original}         & \textbf{Translation} & \textbf{Class} \\\hline
The luminous intensity of the daytime running light must be lower than 400cd. & The luminous intensity of the daytime running light must be LESS THAN 400cd.  & 1 \\\hline
The range of the vehicle shall be at least 450km under WLTP conditions &  The range of the vehicle shall be at least 450km under WLTP conditions.  & 5 \\\hline
If the speed limiter is activated, the vehicle speed shall not exceed the set speed.&  If the speed limiter is activated, the vehicle speed shall not exceed the set speed.& 5 \\\hline
The maximum curb weight of the vehicle must be no more than 3.5t &   The maximum curb weight of the vehicle must be LESS OR EQUAL 3.5t. & 1.\\\hline
The breaking distance can not be longer than 300m.  &   The breaking distance can not be LESS OR EQUAL 300m. & 6 \\\hline
The vehicles horn must not be louder than 50dz. &  The vehicles horn must be LESS OR EQUAL 50dz. &3\\\hline
The vehicles maximum velocity is 260km/h.  &  The vehicles maximum velocity is LESS OR EQUAL 260km/h. &3\\\hline
The minimun distance to a vehicle in front has to be 5m. &   The minimum distance to a vehicle in front has to be LESS OR EQUAL 5m.  &3\\\hline
\end{tabularx}
\label{tab:eval:logic:6}
\end{table}

\begin{table}[h]
\centering
\caption{Evaluation Results: 8 Examples Logical Formulae.}
\begin{tabularx}{0.95\textwidth}{|X|X|l|}
\hline
\textbf{Original}         & \textbf{Translation} & \textbf{Class} \\\hline
The luminous intensity of the daytime running light must be lower than 400cd. &The luminous intensity of the daytime running light must be LESS THAN 400cd.  & 1 \\\hline
The range of the vehicle shall be at least 450km under WLTP conditions & The range of the vehicle shall be at least 450km under WLTP conditions.  & 5 \\\hline
If the speed limiter is activated, the vehicle speed shall not exceed the set speed.&  If the speed limiter is activated, the vehicle speed shall not exceed the set speed.& 5 \\\hline
The maximum curb weight of the vehicle must be no more than 3.5t &  The maximum curb weight of the vehicle must be LESS THAN 3.5t. & 1.\\\hline
The breaking distance can not be longer than 300m.  &  The breaking distance can not be GREATER THAN 300m. & 1 \\\hline
The vehicles horn must not be louder than 50dz. &  The vehicles horn must not be LOUDER THAN 50dz. &5\\\hline
The vehicles maximum velocity is 260km/h.  &   The vehicles maximum velocity is LESS THAN 260km/h. &6\\\hline
The minimun distance to a vehicle in front has to be 5m. &    The minimun distance to a vehicle in front has to be LESS THAN 5m.  &6\\\hline
\end{tabularx}
\label{tab:eval:logic:8}
\end{table}

\begin{table}[h]
\centering
\caption{Support Set: 1 Example If-Then Structure.}
\begin{tabularx}{0.95\textwidth}{|X|X|}
\hline
\textbf{Original}         & \textbf{Translation} \\\hline
If a defective illuminant is detected, the information about the defective illuminant is transmitted to the instrument cluster.     & IF: defective illuminant is detected, THEN: information about the defective illuminant is transmitted to the instrument cluster\\ \hline
\end{tabularx}
\label{tab:supset:ifthen:1}
\end{table}

\begin{table}[h]
\centering
\caption{Support Set: 4 Examples If-Then Structure.}
\begin{tabularx}{0.95\textwidth}{|X|X|}
\hline
\textbf{Original}         & \textbf{Translation} \\\hline
If a defective illuminant is detected, the information about the defective illuminant is transmitted to the instrument cluster.             & IF: defective illuminant is detected, THEN: information about the defective illuminant is transmitted to the instrument cluster\\ \hline
If no advancing vehicle is recognized any more, the high beam illumination is restored within 2 seconds.                                    & IF: no advancing vehicle is recognized any more, THEN: high beam illumination is restored within 2 seconds.\\ \hline
If the light rotary switch is in position "auto", the adaptive high beam headlights are activated by moving the pitman arm to the back.   & IF: the light rotary switch is in position ""auto"" and the pitman arm is moved back, THEN: the adaptive high beam headlights are activated.\\ \hline
(a) When the driver enables the cruise control (by pulling the cruise control lever or by pressing the cruise control lever up or down), the vehicle maintains the set speed if possible.   & IF: driver enables the cruise control by pulling the cruise control lever or by pressing the cruise control lever up or down, THEN: the vehicle maintains the set speed if possible.\\ \hline
\end{tabularx}
\label{tab:supset:ifthen:4}
\end{table}

\begin{table}[h]
\centering
\caption{Support Set: 6 Examples If-Then Structure.}
\begin{tabularx}{0.95\textwidth}{|X|X|}
\hline
\textbf{Original}         & \textbf{Translation} \\\hline
If a defective illuminant is detected, the information about the defective illuminant is transmitted to the instrument cluster.         & IF: defective illuminant is detected, THEN: information about the defective illuminant is transmitted to the instrument cluster\\ \hline
If no advancing vehicle is recognized any more, the high beam illumination is restored within 2 seconds.                                & IF: no advancing vehicle is recognized any more, THEN:high beam illumination is restored within 2 seconds.\\ \hline
If the light rotary switch is in position "auto", the adaptive high beam headlights are activated by moving the pitman arm to the back. & IF: the light rotary switch is in position ""auto"" and the pitman arm is moved back, THEN: the adaptive high beam headlights are activated.\\ \hline
(a) When the driver enables the cruise control (by pulling the cruise control lever or by pressing the cruise control lever up or down), the vehicle maintains the set speed if possible.   & IF: driver enables the cruise control by pulling the cruise control lever or by pressing the cruise control lever up or down, THEN: the vehicle maintains the set speed if possible.\\ \hline
If the driver pushes down the cruise control lever with cruise control activated up to the first resistance level, the speed set point of the cruise control is reduced by N.   & IF: driver pushes down the cruise control lever with cruise control activated up to the first resistance level, THEN: the speed set point of the cruise control is reduced by N.\\ \hline
By pushing the brake or the hand brake, the cruise control is deactivated until it is activated again.  & IF: the brake or the hand brake is pushed, THEN: the cruise control is deactivated until it is activated again.\\ \hline
\end{tabularx}
\label{tab:supset:ifthen:6}
\end{table}

\begin{table}[h]
\centering
\caption{Evaluation Results: 1 Example If-Then Structure.}
\begin{tabularx}{0.95\textwidth}{|X|X|l|}
\hline
\textbf{Original}         & \textbf{Translation} & \textbf{Class} \\\hline
With activated darkness switch (only armored vehicles) the cornering light is not activated.    & IF: darkness switch is activated, THEN: cornering light is not activated  & 3\\ \hline
When moving the pitman arm in position turn left the vehicle flashes all left direction indicators (front left, exterior mirror left, rear left) synchronically with pulse ratio bright to dark 1:1.    & When moving the pitman arm in position turn left the vehicle flashes all left direction indicators (front left, exterior mirror left, rear left) synchronically with pulse ratio bright to dark 1:1.    & 5\\ \hline
If the driver activates the pitman arm during the three flashing cycles of tip-blinking again, only the current flashing cycle is completed.    & IF: the driver activates the pitman arm during the three flashing cycles of tip-blinking, THEN: only the current flashing cycle is completed   & 1\\ \hline
Hazard Warning Light: As long as the hazard warning light switch is released, all direction indicators flash synchronically. If the ignition key is in the ignition lock, the pulse ratio is bright to dark 1:1. If the ignition key is not in the lock, the pulse ratio is 1:2.    &  IF: hazard warning light is released, THEN: all direction indicators flash synchronically. IF: ignition key is in the ignition lock, THEN: pulse ratio is bright to dark 1:1. IF: ignition key is not in the lock, THEN: pulse ratio is 1:2.    & 6\\ \hline
When hazard warning is deactivated and the pit arm is in position direction blinking left or direction blinking right, the direction blinking cycle should be released. & WHEN: hazard warning is deactivated, THEN: direction blinking cycle should be released    & 5\\ \hline
If tip-blinking was activated shortly before deactivation of the hazard warning, this is not considered during the deactivation of the hazard warning   & IF: tip-blinking was activated shortly before deactivation of the hazard warning, THEN: this is not considered during the deactivation of the hazard warning  & 1\\ \hline
(a) The driver activates the low beam headlights by turning the light rotary switch to position exterior lights on  & (a) The driver activates the low beam headlights by turning the light rotary switch to position exterior lights on   & 5\\ \hline
The driver activates the low beam headlights by turning the light rotary switch to position exterior lights on  & The driver activates the low beam headlights by turning the light rotary switch to position exterior lights on    & 5\\ \hline
Distance Warning: The vehicle warns the driver visually and/or acoustically if the vehicle is closer to the car ahead than allowed by the safety distance.  & IF: distance warning is activated, THEN: the vehicle warns the driver visually and/or acoustically    & 6\\ \hline
Emergency Brake Assist: The vehicle decelerates in critical situations to a full standstill & Emergency Brake Assist: The vehicle decelerates in critical situations to a full standstill   & 5\\ \hline
If the driver pushes the cruise control lever up to the first resistance level and the cruise control is activated, the speed set point of the cruise control is increased by the value N.  & IF: the driver pushes the cruise control lever up to the first resistance level, THEN: the speed set point of the cruise control is increased by the value N & 5\\ \hline
\end{tabularx}
\label{tab:eval:ifthen:1}
\end{table}

\begin{table}[h]
\centering
\caption{Evaluation Results: 4 Examples If-Then Structure.}
\begin{tabularx}{0.95\textwidth}{|X|X|l|}
\hline
\textbf{Original}         & \textbf{Translation} & \textbf{Class} \\\hline
With activated darkness switch (only armored vehicles) the cornering light is not activated.    & IF: with activated darkness switch, THEN: the cornering light is not activated.   & 4\\ \hline
When moving the pitman arm in position turn left the vehicle flashes all left direction indicators (front left, exterior mirror left, rear left) synchronically with pulse ratio bright to dark 1:1.    & IF: when moving the pitman arm in position turn left the vehicle flashes all left direction indicators (front left,
exterior mirror left, rear left) synchronically with pulse ratio bright to dark 1:1, THEN: the vehicle flashes all left direction indicators (front left, exterior mirror left, rear left) synchronically with pulse ratio bright to dark 1:1.  & 5\\ \hline
If the driver activates the pitman arm during the three flashing cycles of tip-blinking again, only the current flashing cycle is completed.    & IF: the driver activates the pitman arm during the three flashing cycles of tip-blinking, THEN: only the current flashing cycle is completed. & 1\\ \hline
Hazard Warning Light: As long as the hazard warning light switch is released, all direction indicators flash synchronically. If the ignition key is in the ignition lock, the pulse ratio is bright to dark 1:1. If the ignition key is not in the lock, the pulse ratio is 1:2.    & IF: hazard warning light switch is released, THEN: all direction indicators flash synchronically. If the ignition key is in the ignition lock, the pulse ratio is bright to dark 1:1. If the ignition key is not in the lock, the pulse ratio is 1:2.  & 1\\ \hline
When hazard warning is deactivated and the pit arm is in position direction blinking left or direction blinking right, the direction blinking cycle should be released. & IF: hazard warning is deactivated and the pit arm is in position direction blinking left or direction blinking right, THEN: the direction blinking cycle should be released.   & 1\\ \hline
If tip-blinking was activated shortly before deactivation of the hazard warning, this is not considered during the deactivation of the hazard warning   & IF: tip-blinking was activated shortly before deactivation of the hazard warning, THEN: this is not considered during the deactivation of the hazard warning. & 1\\ \hline
(a) The driver activates the low beam headlights by turning the light rotary switch to position exterior lights on  & IF: the driver activates the low beam headlights by turning the light rotary switch to position exterior lights on, THEN: the low beam headlights are activated.  & 2 \\ \hline
The driver activates the low beam headlights by turning the light rotary switch to position exterior lights on  & IF: the driver activates the low beam headlights by turning the light rotary switch to position exterior lights on, THEN: the low beam headlights are activated.  & 2\\ \hline
Distance Warning: The vehicle warns the driver visually and/or acoustically if the vehicle is closer to the car ahead than allowed by the safety distance.  & IF: distance warning is activated, THEN: the vehicle warns the driver visually and/or acoustically.   & 6 \\ \hline
Emergency Brake Assist: The vehicle decelerates in critical situations to a full standstill & IF: emergency brake assist is activated, THEN: the vehicle decelerates in critical situations to a full standstill    & 6\\ \hline
If the driver pushes the cruise control lever up to the first resistance level and the cruise control is activated, the speed set point of the cruise control is increased by the value N.  & IF: the driver pushes the cruise control lever up to the first resistance level and the cruise control is activated, THEN: the speed set point of the cruise control is increased by the value N.  & 1\\ \hline
\end{tabularx}
\label{tab:eval:ifthen:4}
\end{table}

\begin{table}[h]
\centering
\caption{Evaluation Results: 6 Examples If-Then Structure.}
\begin{tabularx}{0.95\textwidth}{|X|X|l|}
\hline
\textbf{Original}         & \textbf{Translation} & \textbf{Class} \\\hline
With activated darkness switch (only armored vehicles) the cornering light is not activated.    & IF: with activated darkness switch (only armored vehicles), THEN: the cornering light is not activated.   & 2\\ \hline
When moving the pitman arm in position turn left the vehicle flashes all left direction indicators (front left, exterior mirror left, rear left) synchronically with pulse ratio bright to dark 1:1.    & IF: the pitman arm is moved in position turn left, THEN: the vehicle flashes all left direction indicators (front left, exterior mirror left, rear left) synchronically with pulse ratio bright to dark 1:1.  & 1\\ \hline
If the driver activates the pitman arm during the three flashing cycles of tip-blinking again, only the current flashing cycle is completed.    & IF: the driver activates the pitman arm during the three flashing cycles of tip-blinking, THEN: only the current flashing cycle is completed  & 1\\ \hline
Hazard Warning Light: As long as the hazard warning light switch is released, all direction indicators flash synchronically. If the ignition key is in the ignition lock, the pulse ratio is bright to dark 1:1. If the ignition key is not in the lock, the pulse ratio is 1:2.    & IF: the hazard warning light switch is
released, THEN: all direction indicators flash synchronically. If the ignition key is in the ignition lock, the pulse ratio is bright to dark 1:1. If the ignition key is not in the lock, the pulse ratio is 1:2.  & 1\\ \hline
When hazard warning is deactivated and the pit arm is in position direction blinking left or direction blinking right, the direction blinking cycle should be released. & IF: hazard warning is deactivated and the pit arm is in position direction blinking left or direction blinking right, THEN: the direction blinking cycle should be released.   & 1\\ \hline
If tip-blinking was activated shortly before deactivation of the hazard warning, this is not considered during the deactivation of the hazard warning   & IF: tip-blinking was activated shortly before deactivation of the hazard warning, THEN: this is not considered during the deactivation of the hazard warning  & 1\\ \hline
(a) The driver activates the low beam headlights by turning the light rotary switch to position exterior lights on  & IF: the driver activates the low beam headlights by turning the light rotary switch to position exterior lights on, THEN: the low beam headlights are activated.  & 2\\ \hline
The driver activates the low beam headlights by turning the light rotary switch to position exterior lights on  & IF: the driver activates the low beam headlights by turning the light rotary switch to position exterior lights on, THEN: the low beam headlights are activated.  & 2\\ \hline
Distance Warning: The vehicle warns the driver visually and/or acoustically if the vehicle is closer to the car ahead than allowed by the safety distance.  & IF: distance warning is activated, THEN: the vehicle warns the driver visually and/or acoustically.   & 6\\ \hline
Emergency Brake Assist: The vehicle decelerates in critical situations to a full standstill & IF: emergency brake assist is activated, THEN: the vehicle decelerates in critical situations to a full standstill.   & 6\\ \hline
If the driver pushes the cruise control lever up to the first resistance level and the cruise control is activated, the speed set point of the cruise control is increased by the value N.  & IF: the driver pushes the cruise control lever up to the first resistance level and the cruise control is activated, THEN: the speed set point of the cruise control is increased by the value N.  & 1\\ \hline
\end{tabularx}
\label{tab:eval:ifthen:6}
\end{table}

\begin{table}[h]
\centering
\caption{Support Set: 1 Example Modal Verbs.}
\begin{tabularx}{0.95\textwidth}{|X|X|}
\hline
\textbf{Original}         & \textbf{Translation} \\\hline
The maximum deviation of the pulse ratio should be below the cognitive threshold of a human observer.   & The maximum deviation of the pulse ratio MUST be below the cognitive threshold of a human observer.\\ \hline
\end{tabularx}
\label{tab:supset:modal:1}
\end{table}

\begin{table}[h]
\centering
\caption{Support Set: 4 Examples Modal Verbs.}
\begin{tabularx}{0.95\textwidth}{|X|X|}
\hline
\textbf{Original}         & \textbf{Translation} \\\hline
The maximum deviation of the pulse ratio should be below the cognitive threshold of a human observer.   & The maximum deviation of the pulse ratio MUST be below the cognitive threshold of a human observer.\\ \hline
Direction blinking: For USA and CANADA, the daytime running light shall be dimmed by 50\% during direction blinking on the blinking side.   & Direction blinking: For USA and CANADA, the daytime running light MUST be dimmed by 50\% during direction blinking on the blinking side.\\ \hline
The adaptation of the pulse ratio has to occur at the latest after two complete flashing cycles.  & The adaptation of the pulse ratio MUST occur at the latest after two complete flashing cycles.\\ \hline
The duration of a flashing cycle is 1 second.   & The duration of a flashing cycle MUST be 1 second.\\ \hline
\end{tabularx}
\label{tab:supset:modal:4}
\end{table}

\begin{table}[h]
\centering
\caption{Support Set: 6 Examples Modal Verbs.}
\begin{tabularx}{0.95\textwidth}{|X|X|}
\hline
\textbf{Original}         & \textbf{Translation} \\\hline
The maximum deviation of the pulse ratio should be below the cognitive threshold of a human observer. & The maximum deviation of the pulse ratio MUST be below the cognitive threshold of a human observer.\\ \hline
Direction blinking: For USA and CANADA, the daytime running light shall be dimmed by 50\% during direction blinking on the blinking side. & Direction blinking: For USA and CANADA, the daytime running light MUST be dimmed by 50\% during direction blinking on the blinking side.\\ \hline
The adaptation of the pulse ratio has to occur at the latest after two complete flashing cycles.  & The adaptation of the pulse ratio MUST occur at the latest after two complete flashing cycles.\\ \hline
The duration of a flashing cycle is 1 second. & The duration of a flashing cycle MUST be 1 second.\\ \hline
A flashing cycle (bright to dark) will always be completed, before a new flashing cycle can occur. & A flashing cycle (bright to dark) MUST always be completed, before a new flashing cycle can occur.\\ \hline
With subvoltage the ambient light is not available. & With subvoltage the ambient light MUST not be available.\\ \hline
\end{tabularx}
\label{tab:supset:modal:6}
\end{table}

\begin{table}[h]
\centering
\caption{Evaluation Results: 1 Example Modal Verbs.}
\begin{tabularx}{0.95\textwidth}{|X|X|l|}
\hline
\textbf{Original}         & \textbf{Translation} & \textbf{Class} \\\hline
The reaction time between the activation of the pitman arm and the beginning of the first flashing cycle should be below the cognitive threshold of a human observer. & The reaction time between the activation of the pitman arm and the beginning of the first flashing cycle MUST be below the cognitive threshold of a human observer. & 1\\ \hline
The frame rate of the camera is 60 Hz & The frame rate of the camera MUST be 60 Hz & 1\\ \hline
The resolution of the camera signal is 24 bits (8 bits each RGB). & The resolution of the camera signal MUST be 24 bits (8 bits each RGB). & 1\\ \hline
The functions of the system are classified as safety relevant in with respect to ISO 26262.  & The functions of the system are classified as safety relevant in with respect to ISO 26262 & 5\\ \hline
Also after 1000 flashing cycles the cumulated deviation will not exceed 0.05s. & Also after 1000 flashing cycles the cumulated deviation will not exceed 0.05s. & 5\\ \hline
Speed Limit: The vehicle does not exceed a set speed. & The vehicle does not exceed a set speed & 5\\ \hline
A flashing cycle (bright to dark) has to be always completed, before a new flashing cycle can occur. & A flashing cycle (bright to dark) has to be always completed, before a new flashing cycle can occur. & 5\\ \hline
Low beam illuminant shall be LED. & Low beam illuminant shall be LED. & 5\\ \hline
\end{tabularx}
\label{tab:eval:modal:1}
\end{table}

\begin{table}[h]
\centering
\caption{Evaluation Results: 4 Examples Modal Verbs.}
\begin{tabularx}{0.95\textwidth}{|X|X|l|}
\hline
\textbf{Original}         & \textbf{Translation} & \textbf{Class} \\\hline
The reaction time between the activation of the pitman arm and the beginning of the first flashing cycle should be below the cognitive threshold of a human observer. & The reaction time between the activation of the pitman arm and the beginning of the first flashing cycle MUST be below the cognitive threshold of a human observer. & 1\\ \hline
The frame rate of the camera is 60 Hz & The frame rate of the camera MUST be 60 Hz & 1\\ \hline
The resolution of the camera signal is 24 bits (8 bits each RGB). & The resolution of the camera signal MUST be 24 bits (8 bits each RGB). & 1\\ \hline
The functions of the system are classified as safety relevant in with respect to ISO 26262.  & The functions of the system are classified as safety relevant in with respect to ISO 26262. & 5\\ \hline
Also after 1000 flashing cycles the cumulated deviation will not exceed 0.05s. & Also after 1000 flashing cycles the cumulated deviation MUST NOT exceed 0.05s. & 1\\ \hline
Speed Limit: The vehicle does not exceed a set speed. & The vehicle does not exceed a set speed & 5\\ \hline
A flashing cycle (bright to dark) has to be always completed, before a new flashing cycle can occur. & A flashing cycle (bright to dark) MUST be always completed, before a new flashing cycle can occur. & 1\\ \hline
Low beam illuminant shall be LED. & Low beam illuminant shall be LED. & 5\\ \hline
\end{tabularx}
\label{tab:eval:modal:4}
\end{table}

\begin{table}[h]
\centering
\caption{Evaluation Results: 6 Examples Modal Verbs.}
\begin{tabularx}{0.95\textwidth}{|X|X|l|}
\hline
\textbf{Original}         & \textbf{Translation} & \textbf{Class} \\\hline
The reaction time between the activation of the pitman arm and the beginning of the first flashing cycle should be below the cognitive threshold of a human observer. & The reaction time between the activation of the pitman arm and the beginning of the first flashing cycle MUST be below the cognitive threshold of a human observer. & 1\\ \hline
The frame rate of the camera is 60 Hz & The frame rate of the camera MUST be 60 Hz & 1\\ \hline
The resolution of the camera signal is 24 bits (8 bits each RGB). & The resolution of the camera signal MUST be 24 bits (8 bits each RGB). & 1\\ \hline
The functions of the system are classified as safety relevant in with respect to ISO 26262.  & The functions of the system are classified as safety relevant in with respect to ISO 26262. & 5\\ \hline
Also after 1000 flashing cycles the cumulated deviation will not exceed 0.05s. & Also after 1000 flashing cycles the cumulated deviation MUST not exceed 0.05s. & 1\\ \hline
Speed Limit: The vehicle does not exceed a set speed. & The vehicle does not exceed a set speed.  & 5\\ \hline
A flashing cycle (bright to dark) has to be always completed, before a new flashing cycle can occur. & A flashing cycle (bright to dark) MUST always be completed, before a new flashing cycle can occur. & 1\\ \hline
Low beam illuminant shall be LED. & Low beam illuminant MUST be LED.  & 1\\ \hline
\end{tabularx}
\label{tab:eval:modal:6}
\end{table}

\end{document}